%% file: main.tex
\documentclass{article}
\usepackage{spconf,amsmath,amssymb,graphicx}
\usepackage{mathtools}
\usepackage{microtype}
\usepackage{booktabs}
\usepackage[tight-spacing=true]{siunitx}
\usepackage{flushend}
\usepackage{subcaption}
\usepackage{tikz}
\usetikzlibrary{positioning,fit}
\tikzstyle{conn}=[-latex, thick]
\tikzstyle{box}=[draw, rectangle, thick]
\tikzstyle{roundbox}=[box, rounded corners]

\usepackage[backend=biber, bibencoding=utf8,
            style=ieee,
            maxbibnames=99,
            doi=false,
            url=false,
            isbn=false]{biblatex}
\AtEveryBibitem{%
  \ifentrytype{inproceedings}{%
    \clearfield{pages}%
    \clearlist{publisher}%
    \clearlist{organization}%
    \clearlist{location}}{}%
  \ifentrytype{misc}{%
    \clearfield{eprintclass}{} }}

\usepackage[hidelinks]{hyperref}
\usepackage[capitalize]{cleveref}
\crefname{equation}{}{}
\input{macros}

\title{A Flexible Online Framework for Projection-Based STFT Phase Retrieval}
\name{Tal Peer$^{1}$, Simon Welker$^{1,2}$, Johannes Kolhoff$^{\,1}$, Timo Gerkmann$^{1}$\thanks{\scriptsize This work was funded by DASHH (Data Science in Hamburg - HELMHOLTZ Graduate School for the Structure of Matter) --- Grant- No. HIDSS-0002.}}
\address{$^{1}$ Signal Processing (SP), Universität Hamburg, Germany \\
      $^{2}$ Center for Free-Electron Laser Science, DESY, Hamburg, Germany}
\begin{document}
\ninept
\maketitle
\begin{abstract}
Several recent contributions in the field of iterative STFT phase retrieval have demonstrated that the performance of the classical Griffin-Lim method can be considerably improved upon. By using the same projection operators as Griffin-Lim, but combining them in innovative ways, these approaches achieve better results in terms of both reconstruction quality and required number of iterations, while retaining a similar computational complexity per iteration. However, like Griffin-Lim, these algorithms operate in an offline manner and thus require an entire spectrogram as input, which is an unrealistic requirement for many real-world speech communication applications. We propose to extend RTISI---an existing online (frame-by-frame) variant of the Griffin-Lim algorithm---into a flexible framework that enables straightforward online implementation of any algorithm based on iterative projections. We further employ this framework to implement online variants of the fast Griffin-Lim algorithm, the accelerated Griffin-Lim algorithm, and two algorithms from the optics domain. Evaluation results on speech signals show that, similarly to the offline case, these algorithms can achieve a considerable performance gain compared to RTISI.
\end{abstract}
\begin{keywords}
Speech processing, phase retrieval, real-time spectrogram inversion
\end{keywords}
\section{Introduction}
\label{sec:intro}
Many approaches to digital audio signal processing involve operation in the time-frequency domain, most often by application of the short-time Fourier transform (STFT) on the time-domain signal prior to processing. By definition of the discrete Fourier transform (DFT) and in contrast to the real-valued time domain signal, the resulting time-frequency representation is complex-valued and typically expressed in terms of its magnitude and phase components. While some audio processing systems (e.g. for speech enhancement or source separation) operate directly on the complex-valued STFT representation (the \emph{complex spectrogram}), or consider both magnitude and phase components, many systems only take the \emph{magnitude spectrogram} into account. The \emph{phase spectrogram} which is required for the inverse transformation to the time domain is either taken directly from the input signal (if available) or has to be completely estimated from the magnitude spectrogram. The latter case is often referred to as phase retrieval, phase reconstruction, or spectrogram inversion.

Due to the use of finite overlapping frames, the STFT representation is inherently redundant \cite{gerkmannPhaseProcessingSingleChannel2015}. This redundancy gives rise to the notion of \emph{consistency}: A complex spectrogram $\X \in \Cmat{K}{L}$ is said to be consistent if and only if it is in the image of the STFT operation, i.e. if it is the result of an STFT on a time domain signal. If this is not the case, transforming a spectrogram to the time domain and back to the time-frequency domain results in a modified spectrogram $\widetilde{\X} \neq \X$.
In other words, a complex spectrogram $\X$ is only consistent if $\X = \STFT(\iSTFT(\X))$, where $\iSTFT(\cdot)$ denotes the inverse STFT operation \cite{gerkmannPhaseProcessingSingleChannel2015,leroux08b_sapa}.

Iterative projection algorithms are a class of phase retrieval algorithms that make use of consistency by imposing it as a constraint, along with the constraint defined by the known magnitude spectrogram. By iteratively projecting onto sets defined by these two constraints, these algorithms seek an intersection of these two sets: a complex spectrogram that fulfills both constraints by being consistent and having the correct magnitude simultaneously. 
% The problem is non-convex and does not have a unique solution in general. Different algorithms differ in their convergence behavior and in the quality of the solution towards which they converge. 
The basic algorithm used in this context is the Griffin-Lim algorithm (GLA) \cite{griffinSignalEstimationModified1984}, which was proposed in 1984 and is based on earlier works pertaining to phase retrieval in optics. Several extensions of GLA have been proposed through the years, mainly based on ideas and methods from the optimization field \cite{perraudinFastGriffinLimAlgorithm2013,nenovFasterFastAccelerating2023}, borrowing from advances in optics \cite{peerGriffinLimImprovedIterative2022a,kobayashiAcousticApplicationPhase2022} or based on sparsity and other approximations \cite{leroux2010fast}. 

Besides the iterative projection paradigm, other approaches to STFT phase retrieval include model-based methods \cite{beauregardSinglePassSpectrogram2015,masuyamaModelBasedPhaseRecovery2018}, methods based on phase derivatives \cite{prusaNoniterativeMethodReconstruction2017}, optimization techniques \cite{masuyamaGriffinLimPhase2019a} and recently also methods involving deep neural networks (often in combination with one or more of the above) \cite{takamichiPhaseReconstructionAmplitude2018, masuyamaDeepGriffinLim2021, thienInterFrequencyPhaseDifference2023a, peerDiffPhaseGenerativeDiffusionBased2023}. In this work, however, we focus solely on projection-based methods.

Many audio applications call for real-time processing (e.g. speech telecommunications or audio broadcast). Whether or not an algorithm is real-time capable depends on specific implementation and hardware details. Notwithstanding that, a fundamental requirement for any real-time algorithm is online and causal operation: it may only process a relatively small chunk of information at a time and may not rely on future information. Real-Time Iterative Spectrogram Inversion (RTISI), proposed by Beauregard et al. \cite{beauregard2005efficient} and later extended by Zhu et al. \cite{zhuRealTimeIterativeSpectrum2006a,zhuRealTimeSignalEstimation2007}, is an adaptation of GLA to the online setting. Instead of expecting an entire spectrogram as input like GLA and its relatives, RTISI operates on a single frame at a time. By keeping track of the estimated phase of past frames, RTISI is still able to make use of redundancy, although only in the backward direction relative to the current frame. RTISI-LA \cite{zhuRealTimeIterativeSpectrum2006a,zhuRealTimeSignalEstimation2007} relaxes this limitation by introducing ``look-ahead'' frames, i.e. a certain number of future frames that provide redundancy in the forward direction. Although not strictly causal by definition, RTISI-LA can be implemented causally if the added delay complies with the latency requirements of the application at hand. Other extensions to RTISI include modifications to its initialization scheme and order of frame estimation \cite{gnann2010improving}, as well as combinations with other algorithms \cite{leroux2010phase,prusaHighQualityRealTimeSignal2017}.

While other solutions to online STFT phase retrieval have been proposed \cite{pruvsa2016real,masuyamaOnlinePhaseReconstruction2023,binhthienWeightedMisesDistributionbased2023}, RTISI remains a popular choice due to its demonstrated performance and computational efficiency. In this paper, we aim to develop online versions of further iterative algorithms beyond Griffin-Lim. Instead of deriving an online variant for each algorithm specifically, we introduce a more general formulation of RTISI(-LA) that allows quick and straightforward adaptation of any algorithm that is based on the principle of iterative projections.
\section{Iterative Phase Retrieval}\vspace{-5px}
\label{sec:algos}
Iterative projection algorithms transform a given signal by successively enforcing a series of constraints, aiming to converge to a point where all constraints are met. Each constraint is enforced by projecting the signal onto a set defined by that constraint. In the case of STFT phase retrieval, we define two such sets: $\setA$ is the set of all complex spectrograms with a given magnitude spectrogram $\A \in \Rmat{K}{L}$, and $\setC$ is the set of all consistent spectrograms $\X \in \Cmat{K}{L}$. The corresponding projection operations can be defined as
\begin{align}
    \pA(\X) &= \A e^{j \arg(\X)} \eqcomma  \label{eq:pA} \\
    \pC(\X) &= \STFT(\iSTFT(\X) \eqperiod \label{eq:pC}
\end{align}

Iterative projection algorithms operate by applying different combinations of $\pA$ and $\pC$ on the input signal in an iterative manner. Starting from an initialization $\X^0$, we denote the result of iteration $\iter$ as $\X^\iter$. In the following, we briefly present the algorithms that are used in this work. Further details are available in \cite{peerGriffinLimImprovedIterative2022a}.

\subsection{Griffin-Lim algorithm (GLA)}\vspace{-4px}
The classical Griffin-Lim algorithm \cite{griffinSignalEstimationModified1984} simply consists of successive applications of both projections:
\begin{equation}
    \label{eq:GL}
    \X^{\iter+1} = \pC(\pA(\X^\iter)) \eqperiod
\end{equation}

\subsection{Fast Griffin-Lim algorithm (FGLA)}\vspace{-4px}
FGLA \cite{perraudinFastGriffinLimAlgorithm2013} is similar to GLA but includes an auxiliary sequence $\Y$ that is used to increase the step size taken by the algorithm at each iteration, depending on a parameter $\alpha \geq 0$:
\begin{align}
    \Y^{\iter+1} &= \pC(\pA(\X^\iter)) \eqcomma \label{eq:FGL1} \\ 
    \X^{\iter+1} &= \Y^{\iter+1} + \alpha (\Y^{\iter+1} - \Y^\iter) \eqcomma \label{eq:FGL2}
\end{align}
with $\Y^0 = \X^0$. It has been previously shown that FGLA not only converges faster than GLA but may also converge towards a better solution \cite{perraudinFastGriffinLimAlgorithm2013,peerGriffinLimImprovedIterative2022a}. Note that the edge case $\alpha=0$ is equivalent to GLA.

\subsection{Accelerated Griffin-Lim algorithm (AGLA)}\vspace{-4px}
Based on advances in non-convex optimization, AGLA \cite{nenovFasterFastAccelerating2023} is a recently proposed extension of GLA that builds further upon the idea of FGLA and uses two inertial sequences (linearly combined) instead of one:
\begin{align}
    \Y^{\iter+1} &= (1-\gamma) \Z^\iter + \gamma \pC(\pA(\X^\iter)) \eqcomma \\
    \Z^{\iter+1} &= \Y^{\iter+1} + \alpha_1 (\Y^{\iter+1} - \Y^\iter) \eqcomma \\ %alpha_1 -> beta
    \X^{\iter+1} &= \Y^{\iter+1} + \alpha_2 (\Y^{\iter+1} - \Y^\iter) \eqcomma    %alpha_2 -> alpha
\end{align}
with $\alpha_1, \alpha_2, \gamma > 0$ and $\Y^0 = \Z^0 = \X^0$. Although comparatively difficult to tune due to having three parameters, AGLA has been shown to perform better than FGLA for some signals \cite{nenovFasterFastAccelerating2023}. AGLA can be considered a generalization of FGLA: for $\gamma = 1$ the auxiliary variable $\Z^\iter$ is ignored and AGLA reduces to FGLA.

\subsection{Relaxed Averaged Alternating Reflections (RAAR)}\vspace{-4px}
GLA and its extensions above only use the projection operators $\pA$ and $\pC$. It is, however, also possible to define further operators in the space spanned between $\setA$ and $\setC$. The RAAR algorithm, originally proposed for phase retrieval in optics \cite{lukeRelaxedAveragedAlternating2004}, uses the \emph{reflection} about these two sets, defined as
\begin{align}
    \rA(\X) &= 2\pA(\X) - \X \eqcomma \label{eq:rA} \\
    \rC(\X) &= 2\pC(\X) - \X \eqperiod \label{eq:rC} 
\end{align}
Using these reflections, the RAAR algorithm is given by the iteration
\begin{equation}
    \label{eq:RAAR}
    \X^{\iter+1} = \frac{1}{2}\beta \left[ \X^\iter + \rC\left( \rA(\X^\iter) \right) \right] + (1-\beta) \pA(\X^\iter) \eqcomma
\end{equation}
with a relaxation parameter $0 < \beta \leq 1$. Besides its use in optics, RAAR has been shown to perform well for STFT phase retrieval on speech and other audio signals  \cite{kobayashiAcousticApplicationPhase2022,peerGriffinLimImprovedIterative2022a,nenovFasterFastAccelerating2023}.
\subsection{Difference Map (DM)}\vspace{-4px}
Another approach closely related to RAAR is the DM algorithm \cite{elserPhaseRetrievalIterated2003}. DM defines two operations which may be viewed as a generalization of the reflections from \cref{eq:rA,eq:rC}:
\begin{align}
        \fA(\X) &= \pA(\X) + \left( \pA(\X) - \X \right) / \beta \eqcomma \label{eq:DM_fA} \\
        \fC(\X) &= \pC(\X) - \left( \pC(\X) - \X \right)  / \beta  \eqperiod \label{eq:DM_fC} 
\end{align}
The iterative algorithm itself is defined as 
\begin{equation}
    \label{eq:DM}
    \X^{\iter+1} = \X^\iter + \beta \left[ \pC\left(\fA(\X^\iter)\right) - \pA\left(\fC(\X^\iter)\right) \right] \eqcomma
\end{equation}
with relaxation parameter $\beta \neq 0$. As with RAAR, the DM algorithm has also been recently adopted in the context of STFT phase retrieval and has been demonstrated to achieve results superior to both RAAR and FGLA on speech signals \cite{peerGriffinLimImprovedIterative2022a}. Note that for $\abs{\beta} \neq 1$, DM requires computing four projections per iteration, twice as many as all other algorithms discussed here. Furthermore, it can be shown that for $\beta = 1$, RAAR and DM are equivalent.
\section{Online Phase Retrieval}\vspace{-5px}
\label{sec:rtisi}
All algorithms described in \cref{sec:algos} operate directly on an entire spectrogram. While this type of operation is appropriate for some applications, many audio processing systems are designed to work in an online fashion under strict latency requirements. In STFT-based processing, this means that algorithms must operate on a frame-by-frame basis and must be frame-causal, i.e. besides the current frame, an algorithm only has access to past frames. Note that the causality constraint can be slightly violated by artificially delaying the input signal and allowing the algorithm access to a certain number of ``look-ahead'' frames if permitted by latency constraints. 

In order to adapt projection-based algorithms to the online setting, we must first examine how the projections behave in it. The magnitude projection $\pA$ is by definition an element-wise operation and can thus be applied frame-by-frame in a straightforward manner. In contrast, applying the consistency projection $\pC$ on each frame independently would be meaningless, since in this case, the forward and inverse STFTs reduce to simple DFTs and $\pC$ becomes the identity function. In other words, the redundancy which is required for consistency does not exist in this case. However, since an online algorithm does have access to all past frames (and possibly look-ahead frames), one can still make use of the redundancy between those and the current frame and define a modified consistency projection.

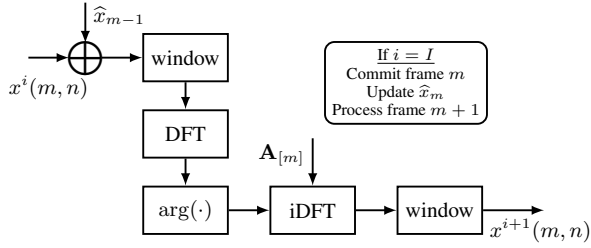
\begin{figure}[t]
    \centering
        \scalebox{0.9}{\input{fig/rtisi.tikz}}
        \caption{Schematic overview of the RTISI algorithm, showing how frame $m$ is processed based on the known magnitude $\A_{\npidx{m}}$ and the windowed past estimate $\widehat{x}_{m-1}$. Each frame is processed for $I$ iterations and then committed by overlap-adding to the past estimate.}
        \label{fig:rtisi}
\end{figure}

The RTISI algorithm \cite{beauregard2005efficient,zhuRealTimeSignalEstimation2007} circumvents this by not explicitly introducing such a projection, but rather defining the algorithm directly on time-domain input and output in terms of overlap-add, windowing, and transforms, as shown in \cref{fig:rtisi}. Frame $m$ of the time domain signal, denoted by $x(m,n)$, is processed iteratively by a windowed DFT, followed by the inverse DFT (iDFT) after replacing the spectral magnitude by the known magnitude for this frame $\A_{\npidx{m}}$. Redundancy is introduced by first overlap-adding $\widehat{x}_{m-1}$, a windowed version of the already-estimated time domain signal based on previous frames. If $B$ look-ahead frames are used, this process is repeated for frames $m,m+1,\dotsc,m+F$ at each iteration.

While this approach leads to a valid (and successful) online version of GLA, it does not allow for simple adaptation of the framework to other iterative algorithms. We propose to reformulate RTISI directly in the time-frequency domain in terms of a frame-wise magnitude projection (identical to $\pA$) and a \emph{partial} consistency projection $\pCp$. By replacing the consistency projection with its partial counterpart, any algorithm based on $\pA$ and $\pC$ can thus be converted into an online algorithm, as opposed to the specially designed nature of RTISI.
\begin{figure}[t]
\centering
        \scalebox{0.9}{\input{fig/framework.tikz}}
        \caption{An online version of GLA implemented using our proposed framework. By rearranging and augmenting the $\pA$ and $\pCp$ blocks, one can easily derive further online algorithms based on the algorithms described in \cref{sec:algos}.}
    \label{fig:rtisi-ng}
\end{figure}
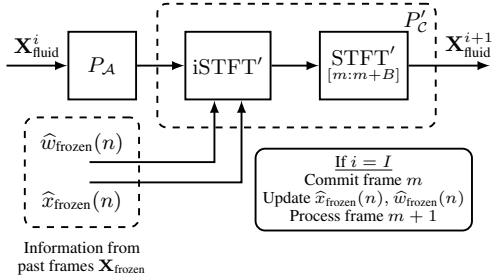
\subsection{Partial consistency projection}\vspace{-3px}
Let us consider a time-domain signal $x(n)$. Its STFT $\X \in \Cmat{K}{L}$ can be defined in terms of its elements (time-frequency bins) $X_{k,\ell}$ as follows: 
\begin{align}
    \label{eq:stft}
    x(n, \ell) &= x(n) w(n-\ell H) \eqcomma \\
    X_{k, \ell} &= \STFT\{x(n)\}_{k,\ell}  = \DFT\{x(n,\ell)\}_k \eqcomma
\end{align}
where $w(n)$ is the window function, $H$ is the frame shift (hop size) and $x(n,\ell)$ are the windowed segments of $x(n)$ before applying the Fourier transform. Assuming the synthesis and analysis windows are identical, the corresponding inverse transform can be defined in the least-squares sense as \cite{griffinSignalEstimationModified1984}
\begin{align}
    \widetilde{x}(n,\ell) &= \iDFT\{X_{k,\ell}\} \eqcomma  \label{eq:istft_seg}\\
    \widetilde{x}(n) &= \iSTFT\{\X\}_n = \frac{\sum\limits_{\ell=0}^{L-1} w(n-\ell H) \widetilde{x}(n,\ell)}{\sum\limits_{\ell=0}^{L-1} w^2(n-\ell H)}  \eqperiod \label{eq:istft}
\end{align}

To define the partial consistency projection we first consider two frame buffers: the frozen buffer contains frames of $\X$ that are committed and will not be changed anymore, while the fluid buffer contains frames that are currently being estimated. Considering the general case with $B$ look-ahead frames, if we have already committed frame $m-1$ and currently processing frame $m$, these buffers can be written as
\begin{align}
    \Xfr &= \X_{\slice{0}{m-1}} =  [X_{k,0}, X_{k,1}, \dotsc, X_{k,m-1}] \eqcomma \\
    \Xfl &= \X_{\slice{m}{m+B}} = [X_{k,m}, X_{k,m+1}, \dotsc, X_{k,m+B}] \eqperiod
\end{align}
Using these buffers we can define a partial iSTFT by splitting the numerator and denominator of \cref{eq:istft} into frozen and fluid parts:
\begin{equation}
    \label{eq:istft_partial}
    \begin{split}
    \iSTFT'\{&\Xfl \mid \Xfr\}_{n} = \\ 
    &\frac{\overbrace{\textstyle \sum\limits_{\ell=0}^{m-1} w(n-\ell H) \widetilde{x}(n,\ell)}^{\SumXfr} + \sum\limits_{\ell=m}^{m+B-1} w(n-\ell H)\widetilde{x}(n,\ell)}
    {\underbrace{\textstyle \sum\limits_{\ell=0}^{m-1} w^2(n-\ell H)}_{\SumWfr} + \sum\limits_{\ell=m}^{m+B-1} w^2(n-\ell H)} \eqperiod
    \end{split}
\end{equation}
Since frozen frames will not be changed anymore, it is not necessary to compute the iDFT for them, and the partial windowed signal $\SumXfr$ and squared window sum $\SumWfr$ contain all information regarding the frozen frame buffer. These sums can be tracked and updated after each frame is committed, to be used for the next frame, and so on. Furthermore, since each frame only overlaps with a finite number of past frames, an efficient implementation of \cref{eq:istft_partial} should only compute samples that belong to the part of $\Xfl$ that overlaps with $\Xfr$. With frame length $N$ and frame shift $H$, the first sample to compute for frame $m$ is positioned at $n_0 = H \left(m - \left\lceil \frac{N}{H} -1 \right\rceil \right)$.

Finally, the partial projection of the fluid buffer $\Xfl$ given a frozen buffer $\Xfr$ is defined analogously to \cref{eq:pC}:
\begin{equation}
    \label{eq:pC_partial}
    \pCp(\Xfl \mid \Xfr) = \STFTp_{\slice{m}{m+B}} \{\iSTFT'\{\Xfl\mid \Xfr\}\} \eqperiod
\end{equation}
where STFT$'$ denotes an STFT operation that is only applied to the frames belonging to $\Xfl$, such that older frames are not altered (for $B=0$ this reduces into a windowed DFT). As illustrated in \cref{fig:rtisi-ng}, RTISI(-LA) can be simply defined as the online variant of GLA by adjusting \cref{eq:GL} to use $\pCp$:
\begin{equation}
    \Xfl^{\iter+1} = \pCp(\pA(\Xfl^\iter) \mid \Xfr) \eqperiod
\end{equation}
Similarly, all other algorithms from \cref{sec:algos}, as well as any other projection-based STFT phase retrieval algorithm, can be converted to online operation in the same manner. This also applies to deep learning methods that are based on the iterative projection paradigm, such as the Deep Griffin-Lim Iteration (DeGLI) \cite{masuyamaDeepGriffinLim2021}.
\begin{figure}
    \centering
    \includegraphics[width=0.95\columnwidth]{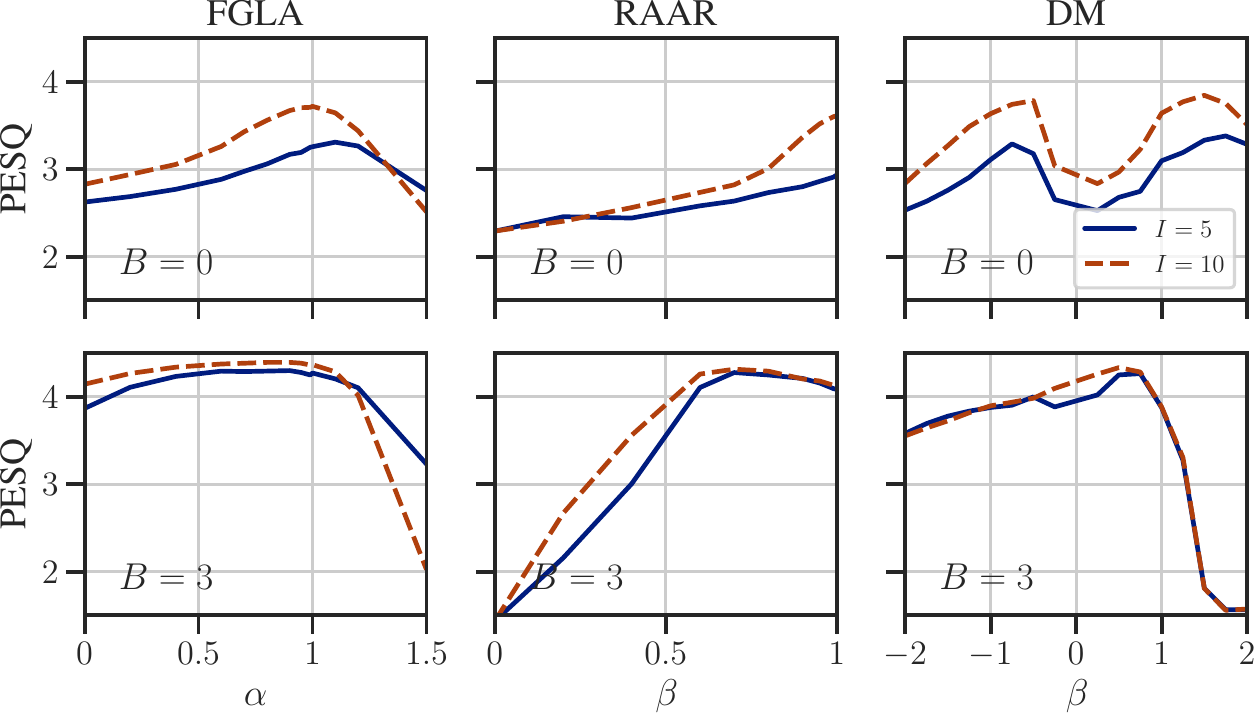}
    \caption{Mean PESQ score as a function of $\beta$ for different algorithms.}
    \label{fig:param_search}
\end{figure}
\begin{table}[t]
\centering
\scalebox{0.95}{
\begin{tabular}{@{}lccccc@{}}
\toprule
        && $B=0$         && $B>0$         &\\ \midrule
FGLA    && $\alpha=0.99$ && $\alpha=0.8$  &\\
RAAR    && $\beta=0.99$   && $\beta=0.7$  &\\
DM      && $\beta=1.5$   && $\beta=0.5$   &\\ \midrule
AGLA    && \multicolumn{3}{c}{$\alpha_1=0.95, \alpha_2=0.99, \gamma=1.2$} &\\ \bottomrule
\end{tabular}
}
\caption{Best performing parameter values with and without look-ahead. Values for FGLA, RAAR, and DM are based on our parameter search (\cref{fig:param_search}). Values for AGLA are taken from \cite{nenovFasterFastAccelerating2023}.}\label{tab:best_params}
\end{table}
\section{Experimental Evaluation}\vspace{-5px}
\label{sec:experiments}
We evaluate the proposed framework using GLA (equivalent to RTISI) and the other algorithms from \cref{sec:algos} as underlying algorithms, with various numbers of iterations $\imax$ and number of look-ahead frames $B$. For FGLA, DM, and RAAR, which have a single tuning parameter, we first perform a parameter search over a plausible range for each algorithm. In the case of AGLA we simply choose the parameter values that are described as best on average in \cite{nenovFasterFastAccelerating2023}. In all cases we use the same phase initialization scheme as RTISI: after the first $B+1$ frames are initialized with zero-phase, each new frame in $\Xfl$ is initialized by taking the STFT of the signal estimate up to this point including the phaseless new frame.

\subsection{Experimental setting and metrics}
All experiments are performed on 100 spoken utterances (gender-balanced) from the TIMIT corpus \cite{timit}, sampled at \qty{16}{\kilo\hertz}. All signals are transformed to the STFT domain using a frame length of \qty{32}{\ms}, a frameshift of \qty{8}{\ms}, and a Hann window function. For each utterance $\mathbf{S}$, the magnitude spectrogram $\A = \abs{\mathbf{S}}$ is kept and the phase spectrogram is replaced by zero. 

We use the PESQ score (Perceptual Evaluation of Speech Quality) to assess the perceptual quality of reconstructed signals. We also report the spectral convergence (SC), which is a non-perceptual error metric common in phase retrieval studies \cite{sturmel2011signal,masuyamaDeepGriffinLim2021}.

\subsection{Results}\vspace{-3px}
\label{sec:results}
\begin{figure}
    \centering
    \includegraphics[width=\columnwidth]{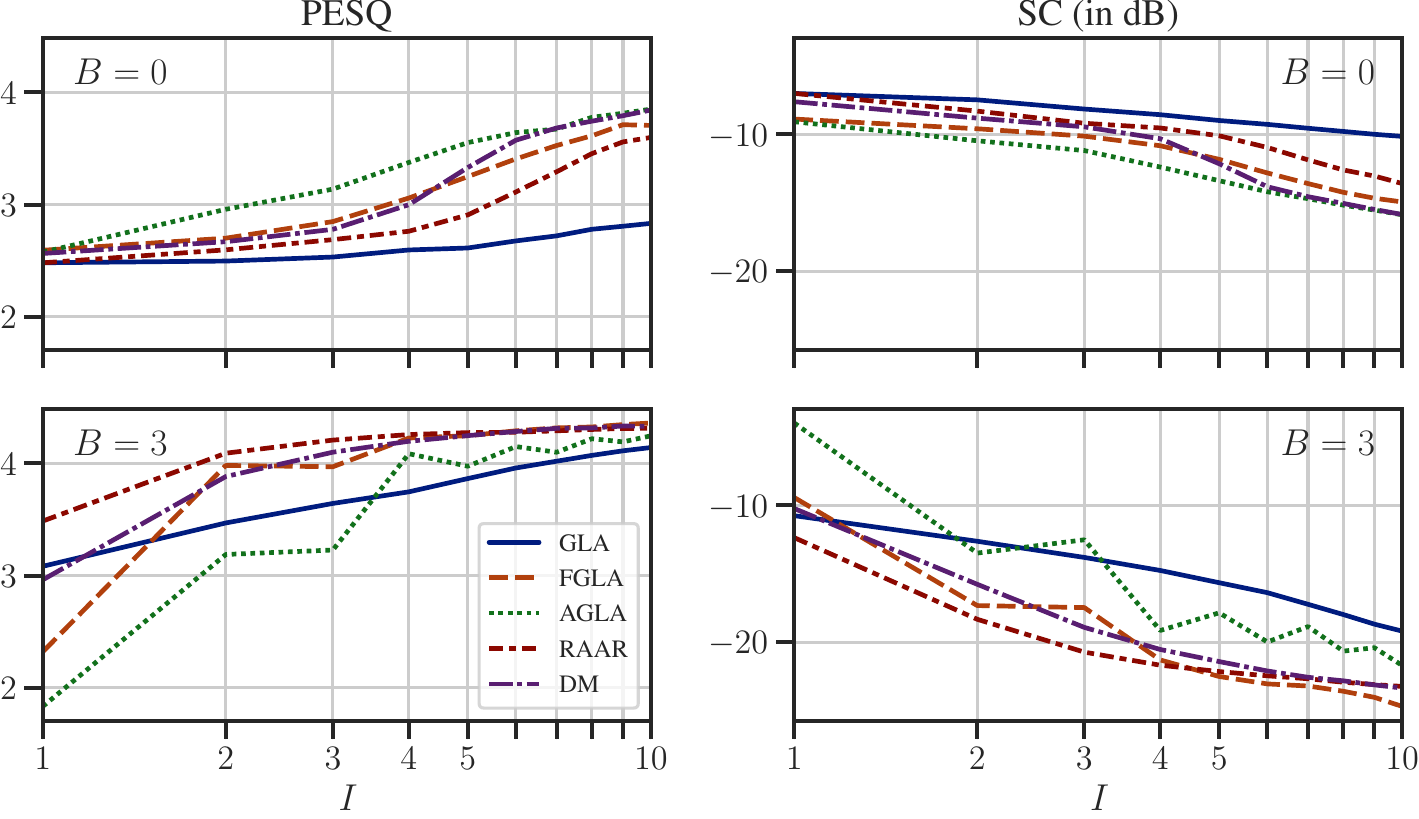}
    \caption{Mean PESQ and SC as a function of the number of iterations $\imax$. Except for GLA, each algorithm is evaluated using the parameter values given in \cref{tab:best_params}. The horizontal axis is scaled logarithmically.}
    \label{fig:pesq_sc}
\end{figure}
Results of our parameter search are shown in \cref{fig:param_search} in terms of reconstruction quality (mean PESQ) for each algorithm. For brevity, we only show results for $B \in \{0,3\}$, but we note that the behavior for other values of $B>0$ is largely similar to $B=3$. DM shows a comparatively strong sensitivity to the choice of $\beta$, while FGLA delivers good performance across a wide range of $\alpha$, especially for $B>0$. For all three algorithms, the optimal parameter value differs considerably between the look-ahead and non-look-ahead cases. Based on this observation, we use different parameter values for these two cases in the following. The chosen values are listed in \cref{tab:best_params}, along with the optimal parameter values for AGLA as determined in \cite{nenovFasterFastAccelerating2023}. Note that this parameter search is specific to the STFT parameters and type of signals (speech) that we use.

Having found good parameter values, we now analyze the behavior of all algorithms for different numbers of iterations. In \cref{fig:pesq_sc} we report results for $\imax \in \{1,2,\dotsc,10\}$ in terms of PESQ and SC. Compared to GLA (equivalent to RTISI), all algorithms perform considerably better using $\imax = 10$ iterations per frame. This is consistent with results for the offline case \cite{perraudinFastGriffinLimAlgorithm2013,peerGriffinLimImprovedIterative2022a,nenovFasterFastAccelerating2023}. Focusing on the low-iteration regime, we again observe a difference with respect to the number of look-ahead frames $B$. For $B=0$, AGLA consistently performs well, while FGLA, RAAR, and DM only begin to perform considerably better than GLA for $\imax> 3$. However, when including look-ahead frames, RAAR and DM are able to achieve excellent reconstruction quality already with a single iteration, while FGLA and AGLA need a few more iterations to catch up. The same trend has been observed for other values of $B>0$. With parameter tuning, AGLA might perform better for the look-ahead case, but we do not pursue this here. Especially impressive is RAAR's ability to achieve a PESQ score of about 3.5 with a single iteration per frame and a computational complexity similar to that of GLA.
\section{Conclusion}\vspace{-4px}
\label{sec:conclusion}
This paper proposes a generalization of the RTISI method for online STFT phase retrieval. By introducing a modified consistency projection and reformulating RTISI in terms of this projection, we are able to show that RTISI, although specifically based on the Griffin-Lim algorithm, can be extended beyond the simple GLA iteration and be used to implement online variants of any iterative projection algorithm in a simple and effective way. The new framework is validated by evaluating on a dataset of speech signals using various underlying algorithms. Evaluation results show that the performance boost offered by improved offline STFT phase retrieval algorithms is also evident in the online case and results in a considerable improvement over RTISI(-LA). The proposed framework will also enable easy online adaptation of novel projection-based algorithms introduced in the future.
\clearpage
\section{References}
\label{sec:refs}
\atColsBreak{\vskip5pt}
\printbibliography[heading=none]
\end{document}

%% file: macros.tex
\usepackage{amssymb,amsmath}
\usepackage{mathtools}

\newcommand{\setA}{\mathcal{A}}
\newcommand{\setC}{\mathcal{C}}

\newcommand{\pA}[0]{P_\setA}
\newcommand{\pC}[0]{P_\setC}
\newcommand{\pCp}[0]{P'_\setC}

\newcommand{\rA}[0]{R_\setA}
\newcommand{\rC}[0]{R_\setC}

\newcommand{\fA}[0]{f_\setA}
\newcommand{\fC}[0]{f_\setC}

\newcommand{\mat}[1]{\mathbf{#1}}

\newcommand{\A}[0]{\mat{A}}
\newcommand{\X}[0]{\mat{X}}
\newcommand{\Y}[0]{\mat{Y}}

\newcommand{\Z}[0]{\mat{Z}}

\newcommand{\Xfr}[0]{\X_\text{frozen}}
\newcommand{\Xfl}[0]{\X_\text{fluid}}

\newcommand{\SumWfr}[0]{\widehat{w}_{\text{frozen}}(n)}
\newcommand{\SumXfr}[0]{\widehat{x}_{\text{frozen}}(n)}

\newcommand{\iter}[0]{i}
\newcommand{\imax}[0]{I}

\newcommand{\npidx}[1]{[#1]}
\newcommand{\slice}[2]{\npidx{#1:#2}}

\DeclareMathOperator{\STFT}{STFT}
\DeclareMathOperator*{\STFTp}{STFT^{\mathrm{\prime}}}
\DeclareMathOperator{\iSTFT}{iSTFT}

\DeclareMathOperator{\DFT}{DFT}
\DeclareMathOperator{\iDFT}{iDFT}
\newcommand{\eqcomma}{\,,} % comma after equation
\newcommand{\eqperiod}{\,.} % period after equation
\newcommand{\minitimes}{{\mkern-2mu\times\mkern-2mu}}

\DeclarePairedDelimiter{\abs}{\lvert}{\rvert}

\newcommand{\R}{\mathbb{R}}
\newcommand{\C}{\mathbb{C}}
\newcommand{\Rmat}[2]{\R^{#1\minitimes#2}}
\newcommand{\Cmat}[2]{\C^{#1\minitimes#2}}

%% file: fig/rtisi.tikz
\begin{tikzpicture}[node distance=0.4cm and 0.6cm]
    \tikzstyle{nicebox}=[box, minimum width=1.25cm, minimum height=0.7cm]
    \node (x_in) {};
    \node[right=of x_in, inner sep=0, outer sep=-0.4mm] (add) {\Large $\bigoplus$};
    \node[nicebox, right=of add] (w1) {window};
    \node[nicebox, below=of w1] (dft) {DFT};
    \node[nicebox, below=of dft] (arg) {$\arg(\cdot)$};
    \node[nicebox, right=of arg] (idft) {iDFT};
    \node[nicebox, right=of idft] (w2) {window};
    \node[right=of w2, xshift=3mm] (x_out) {};

    \node[above=of add,yshift=2mm] (x_old_in) {};
    \node[above=of idft, yshift=3mm] (A_in) {};

    \draw[conn] (x_in) -- node[below,,yshift=-1.5mm] {$x^i(m,n)$} (add);
    \draw[conn] (add) -- (w1);
    \draw[conn] (w1) -- (dft);
    \draw[conn] (dft) -- (arg);
    \draw[conn] (arg) -- (idft);
    \draw[conn] (idft) -- (w2);
    \draw[conn] (w2) -- node[below, xshift=4mm] {$x^{i+1}(m,n)$} (x_out);

    \draw[conn] (x_old_in) -- node[right,yshift=1mm] {$\widehat{x}_{m-1}$} (add);
    \draw[conn] (A_in) -- node[left,yshift=1mm] {$\A_{\npidx{m}}$} (idft);

    \node[roundbox, above=of w2, yshift=5mm, xshift=-5mm, align=center, font=\scriptsize] (next_frame) {\underline{If $i=\imax$}\\Commit frame $m$\\Update $\widehat{x}_{m}$\\Process frame $m+1$};
            % \node (X_fl_in) {};
            % \node[box, minimum height=1cm, minimum width=1cm, right=of X_fl_in, xshift=0.2cm] (pA) {$\pA$};
            % \node[box, minimum height=1cm, right=of pA] (iSTFT) {$\iSTFT'$};
            % \node[box, minimum height=1cm, right=of iSTFT] (STFT) {$\displaystyle \STFTp_{[m:m+F]}$};
            % \node[right=of STFT, xshift=0.5cm] (X_fl_out) {};

            % \node[below=of pA, yshift=-0.6cm, xshift=-0.3cm] (Sw_in) {}; 
            % \node[below=of pA, yshift=-0.9cm, xshift=-0.3cm] (Sx_in) {}; 

            % \node[roundbox, below=of STFT, yshift=-0.5cm, align=center, font=\scriptsize] (next_frame) {\underline{If $i=\imax$}\\Commit frame $m$\\Update $\SumXfrM,\SumWfrM$\\Process frame $m+1$};

            % \draw[conn] (X_fl_in) -- node [above] {$\Xfl^\iter$} (pA);
            % \draw[conn] (pA.east) -- (iSTFT);
            % \draw[conn] (iSTFT.east) -- (STFT);
            % \draw[conn] (STFT) -- node [above,xshift=3mm] {$\Xfl^{\iter+1}$} (X_fl_out);

            % \draw[conn] (Sw_in) node[above] (Sw_lbl) {$\SumWfr$} -| ([xshift=-0.2cm]iSTFT.south);
            % \draw[conn] (Sx_in) node[below] (Sx_lbl) {$\SumXfr$} -| ([xshift=0.2cm]iSTFT.south);

            % \node[roundbox, dashed, fit=(iSTFT)(STFT), inner sep=4mm] (pC_box) {};
            % \node[text=black, left=0cm of pC_box.north east, anchor=north east] {$\pCp$};
            
            % \node[roundbox, dashed, fit=(Sw_lbl)(Sx_lbl)] (past_box) {};
            % \node[below=of past_box, align=center,font=\scriptsize, outer sep=0, inner sep=0] {Information from\\past frames $\Xfr$};
            
\end{tikzpicture}

%% file: fig/framework.tikz
        \begin{tikzpicture}[node distance=0.2cm and 0.7cm]
            \node (X_fl_in) {};
            \node[box, minimum height=1cm, minimum width=1cm, right=of X_fl_in, xshift=0.2cm] (pA) {$\pA$};
            \node[box, minimum height=1cm, right=of pA] (iSTFT) {$\iSTFT'$};
            \node[box, minimum height=1cm, right=of iSTFT] (STFT) {$\displaystyle \STFTp_{[m:m+B]}$};
            \node[right=of STFT, xshift=0.5cm] (X_fl_out) {};

            \node[below=of pA, yshift=-0.6cm, xshift=-0.3cm] (Sw_in) {}; 
            \node[below=of pA, yshift=-0.9cm, xshift=-0.3cm] (Sx_in) {}; 

            \node[roundbox, below=of STFT, yshift=-0.5cm, align=center, font=\scriptsize] (next_frame) {\underline{If $i=\imax$}\\Commit frame $m$\\Update $\SumXfr,\SumWfr$\\Process frame $m+1$};

            \draw[conn] (X_fl_in) -- node [above] {$\Xfl^\iter$} (pA);
            \draw[conn] (pA.east) -- (iSTFT);
            \draw[conn] (iSTFT.east) -- (STFT);
            \draw[conn] (STFT) -- node [above,xshift=3mm] {$\Xfl^{\iter+1}$} (X_fl_out);

            \draw[conn] (Sw_in) node[above] (Sw_lbl) {$\SumWfr$} -| ([xshift=-0.2cm]iSTFT.south);
            \draw[conn] (Sx_in) node[below] (Sx_lbl) {$\SumXfr$} -| ([xshift=0.2cm]iSTFT.south);

            \node[roundbox, dashed, fit=(iSTFT)(STFT), inner sep=4mm] (pC_box) {};
            \node[text=black, left=0cm of pC_box.north east, anchor=north east] {$\pCp$};
            
            \node[roundbox, dashed, fit=(Sw_lbl)(Sx_lbl)] (past_box) {};
            \node[below=of past_box, align=center,font=\scriptsize, outer sep=0, inner sep=0] {Information from\\past frames $\Xfr$};
            
        \end{tikzpicture}